# Thermomechanical Processing of Pure Magnesium: Hot Extrusion, Hot Rolling and Cold Drawing

M. A. Kalateh[1], N. Talebi[1], S. Nekoei[1], M. M. Novini[1], F. Khodabakhshi[1], M. Nili Ahmadabadi[1*]

[1] School of Metallurgy and Materials Engineering, College of Engineering, University of Tehran, Tehran, Iran

*(corresponding author email address: nili@ut.ac.ir)

*Abstract-A comprehensive study on thermomechanical processing of pure Mg was conducted through sequential hot extrusion, hot rolling, and cold drawing operations. Three different extrusion ratios (6:1, 25:1, and 39:1) were investigated at 350°C, revealing that 39:1 ratio produced an optimal bimodal grain structure with beneficial twin morphology. Subsequently, hot rolling experiments were performed at varying linear speeds (26- and 130-mm s$^{-1}$) and interpass annealing times (2.5 and 10 minutes). Results demonstrated that higher rolling speeds led to finer microstructure, while longer interpass annealing times resulted in reduced twin fraction and more inhomogeneous microstructure. The processed material was then subjected to cold drawing with approximately 12% true strain per pass. Different annealing conditions (275°C and 375°C for 2.5-10 minutes) between drawing passes were evaluated. Analysis showed that annealing at 375°C for 2.5-5 minutes provided optimal softening for subsequent deformation. Fracture analysis revealed a mixed ductile-brittle behavior, with twin-matrix interfaces serving as preferred crack propagation paths This study establishes optimal processing parameters for pure Mg wire production, highlighting the critical role of twin characteristics and restoration processes in determining material formability during multi-step thermomechanical processing.*

## I. Introduction

Owing to the global concern about energy consumption, special considerations have taken place on research and development of lightweight alloys [1]. Among lightweight alloys, Mg has gained a stunning attention due to its low density (1.74 g/cm$^3$) and as a result, its high specific strength (130 kNm/kg). Also, some exclusive characteristics of Mg, such as high volumetric capacity for storing Hydrogen and excellent biocompatibility have made it an inevitable candidate for Energy and medical applications [2-4]. However, owing to the high tendency of Pure Mg and its alloys to form a strong basal texture during cold (in some cases, even hot) deformation, there's a huge drawback in production of Mg parts.

It is believed that the main reason behind the weak formability of Mg is its HCP crystal structure with insufficient active slip systems at room temperature. In fact, basal slip dominates the room temperature deformation of Mg (due to the lowest critical resolved shear stress) and the resulting unfavored basal texture, sacrifices ductility and inhibits further straining. Also, several reports have mentioned that improper thermomechanical processing routes and heat treatments would strengthen the undesirable basal <a> texture and limit further formability.

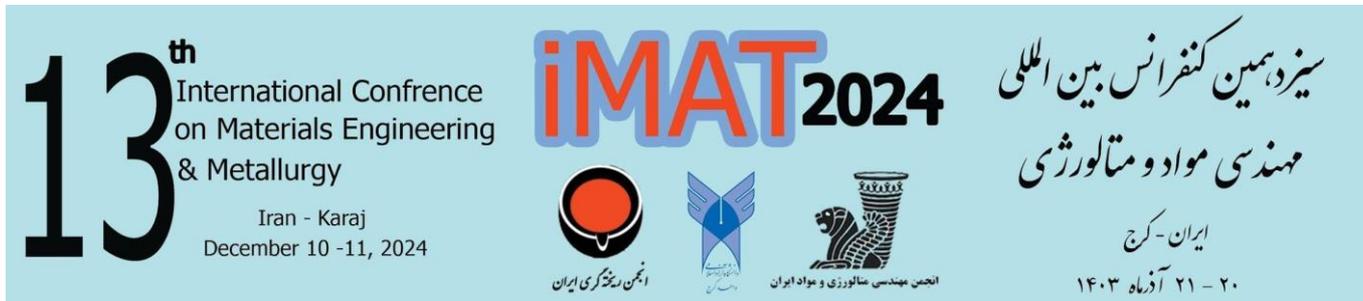

To overcome this lack of strain accommodation in Mg and its alloys, the main attempts could be categorized as Texture modification [5], alloying with rare-earth elements to optimize the c/a ratio and formation of RE-texture [2] and last but not least, grain refinement [6]. Among the various methods, grain refinement through recrystallization enhances ductility while simultaneously reducing texture intensity, resulting in a more randomized texture due to the formation of new, strain-free grains. In Mg alloys, controlling texture development and achieving grain refinement are often accomplished through recrystallization. Static recrystallization (SRX) plays a critical role in weakening the texture, a process influenced by the types of active dislocations and twins [7]. Additionally, previous studies have shown that SRX is accompanied by the activation of non-basal slip and twinning [8]. Dynamic recrystallization (DRX), on the other hand, occurs readily in magnesium (Mg) during thermomechanical processing due to its low melting point, which usually happens upon practical hot deformation processes such as rolling, forging and extrusion. Although Mg has a relatively high stacking fault energy (SFE), which would typically limit DRX during hot deformation, the limited number of slip systems in its hexagonal close-packed (HCP) structure makes DRX a prominent grain-refinement mechanism in Mg alloys, especially during elevated-temperature thermomechanical processing when twinning is suppressed. Studies report that several DRX mechanisms—including discontinuous DRX (DDRX), continuous DRX (CDRX), and twinning-induced DRX (TDRX)—commonly occur in wrought Mg alloys. Many factors, such as strain rate, deformation temperature, applied strain, and alloying elements, influence both twinning and DRX activation [9].

Findings reveal that an effective way to control grain refinement through dynamic recrystallization (DRX) in Mg alloys is by adjusting the Zener-Hollomon parameter ($Z = \dot{\varepsilon} \exp(Q/RT)$), which encapsulates both deformation temperature and strain rate. According to this relationship, increasing the strain rate and lowering the deformation temperature enhance the fraction of dynamically recrystallized grains [9]. Extrusion Parameters, such as temperature, strain rate and extrusion ratio (ER, i.e., Degree of Deformation) effectively determine the fraction and size of the DRXed grains, and characteristic of twins. Previous studies have proved that the extrusion ratio is the most important parameter in the whole extrusion process that can tune the grain size distribution and resulting texture [10]. Herein, Chen et al. [11] have shown that for MG AZ31 alloy, ER≥39 AND ER≤24 will be more and less effective on grain refinement, respectively. However, alloying elements will change the activation energy for restoration processes, which means these presented criteria may not be reputable for CP Mg, but the general concept is true.

In this work, various thermomechanical routes (Hot extrusion, Hot rolling and Cold drawing) were performed on commercially pure Mg to shape a continuous wire with desirable performance for additive manufacturing of this material. The main reason behind selection of commercially pure Mg, is to avoid the strong effect of solute atoms on microstructural evolution, such as grain refinement and texture modification, and investigate the direct effect of thermomechanical processing parameters on its microstructure and mechanical behavior.

## II. MATERIALS AND METHODS

**Thermomechanical Procedure:**

**Hot Extrusion:** In this study, Commercially Pure (CP) as-cast Mg ingot was utilized. The as-cast ingot was sliced into ɸ25*100 mm cylinders using an electro-discharge machine and after surface cleaning by 180 grit SiC sandpaper, cylindrical rods were subjected to hot extrusion. Extrusion experiments were performed at 350 °C with a strain rate of 0.04 s$^{-1}$ and three extrusion ratios were considered, which are 6:1 (ɸ$_f$=10mm), 25:1 (ɸ$_f$=5 mm) and 39:1 (ɸ$_f$=4 mm). The cylinders were heated up to extrusion temperature with a 10-minute soaking time in a resistant furnace. In addition, whole dies and containers were also heated up to the extrusion temperature. It is worth mentioning that the deformation cone was made of H13 tool steel with a die half angle of 25 degree.

**Hot Rolling:** As-extruded products were subjected to Hot Rolling at 350 °C. Two different rolling speeds (26- and 130-mm s$^{-1}$) were considered in order to investigate the effect of rolling rate on the microstructural evolution. The applied logarithmic strain in each pass of hot rolling was about 40 percent, and the soaking time between each pass of hot rolling was considered as a variable in this series of experiments, with values of 2.5 and 10 minutes. The rolling strategy was considered as "one-way hot rolling", which means that the deformation in all stages was starting from the initial nib of the rod.

**Cold Drawing:** Typically, wire drawing is used in order to reach a smoother surface, more homogeneous deformation and lower diameters. So, as-rolled samples (ɸ 2 mm) were subjected to multi-pass cold wire drawing. Due to high susceptibility of pure Mg to wire breakage during cold drawing, many attempts have taken place to avoid failure of the drawing process [12]. It has been reported that there are two key factors affecting selection of proper dies for wire drawing of pure Mg, which are die half angle & reduction per pass. Taking into account, tungsten carbide dies with a half angle of 6 degree and reduction per pass of 10-13% in cross section were chosen, based on [12, 13]. Also, for decreasing the friction coefficient and control redundant work during wire drawing, a highly viscous oil-based lubricant was used to wet the surface of wire, while passing the deformation zone of the die.

As aforementioned, wire breakage during cold drawing of pure Mg is of its main drawbacks, which could be avoided by the proper annealing between passes. Due to High stacking fault energy (125 mJ m$^{-2}$) and low melting point (660 °C) of pure magnesium, dislocation motion and interdependent phenomena such as recovery and recrystallization are more sensitive to the temperature range and applied strain. Since the applied strain during wire drawing is kept constant, so investigation of proper annealing condition for reviving ductility and drawability took place by varying annealing time (2.5, 5, 7.5 and 10 minutes) and temperature (275 and 375 °C).

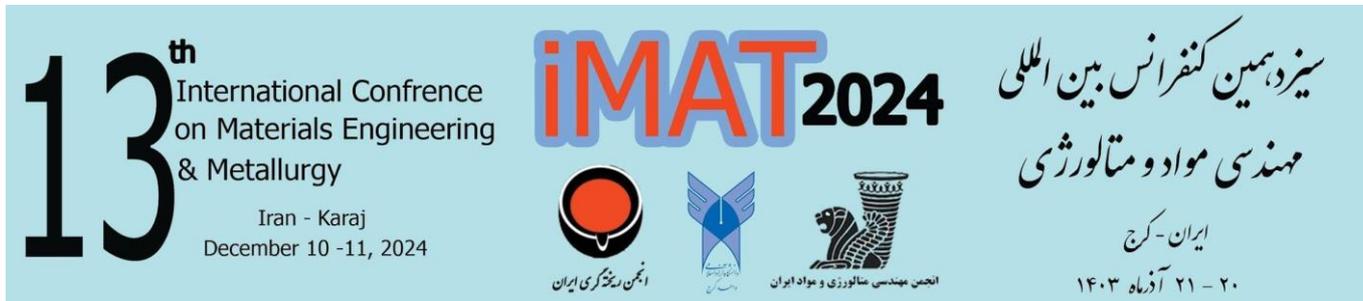

**Microstructural Characterizations:**

Cross- & transverse-sections of the as-extruded specimens were cut and then mechanically polished until achieving a mirror-finished specimen. Subsequently, the samples were chemically etched with an etchant of 4.2 g picric acid, 10 ml acetic acid, 10 ml distilled water and 70 ml of ethanol for optical microscopy observations (OM, Zeiss Axioskop 2). The fractured surface of the wires which were broken upon cold drawing was studied through Secondary Electron images resulting from SEM. The SEM images were captured using Philips XL30S-FEG at 25kV. Also, image analyses were performed using ImageJ software. Vickers microhardness measurements were performed at the center of the longitudinal section perpendicular to the deformation direction. To provide this, a load of 100 g was applied for a dwell time of 10s, and the mean microhardness values were obtained from the average of seven separate hardness measurements. It is worth mentioning that in order to avoid hardness trace overlapping, the 0.3 mm spatial interval between each measurement was noticed.

## III. RESULTS AND DISCUSSION

**Hot Extrusion:** Fig. 1 depicts the microstructure of the as-extruded samples at different extrusion ratios, supported by the statistical analyses of the grain size distribution in Table I. Comparison between ER 6:1 and ER 25:1, declares that the higher extrusion ratio has led to higher fraction of smaller recrystallized grains, as a result of higher strain and the superior amount of stored energy. This case can be validated by the lower coefficient of variation (CV) value of the ER 25:1. Also, the lower value of CV indicates better homogeneity of the hot deformation, which is aligned with the previous investigations [14].

ER 39:1 shows a different trend, with higher deformation leading to coarser grains (Table I). The high CV value and grain size distribution indicate a bimodal structure, with two major grain groups: SGs (10-20 µm) and LGs (60-70 µm). It can be considered that SGs are as a result of DRX, and this hypothesis is aligned with the fact that higher extrusion ratios will lead to smaller DRXed grains. Discussion about LGs needs the accurate investigation of Fig. 1 (c1). Imprimis, there are subgrains all over the microstructure with the size in the scale of less than 10 µm. Secondly, in the LGs there are ultra-thin twin lamellae, which coexist with these subgrains. The hypothesis suggests that the large strain at ER 39:1 provides sufficient energy to promote DRX, resulting in smaller DRXed grains, observed as SGs in the histogram. Meanwhile, Fig. 1(c1) shows LG regions in the microstructure with uniformly distributed ultra-thin twin lamellae inside the grains. The LGs in ER 39:1 likely result from high adiabatic heat generated during large-strain thermomechanical processing. The ultra-thin twin lamellae likely form repetitively to accommodate the high strain energy. For further discussion, the twin area fraction is reported in Table I.

At ER 6:1, a high twin area fraction (46.62%) is observed, attributed to easier twinning in coarser grains (35.71 µm). In contrast, ER 25:1 shows less twinning due to a higher fraction of DRXed grains and strain energy consumption through DRX. However, ER 39:1 deviates the trend, showing a higher twin fraction than ER 25:1, confirming that adiabatic heat promotes coarser grains and periodic ultra-thin twin lamellae to accommodate high strain energy. The twin morphology, including length, width, and thickness, changes with ER. At ER 6:1, twins are thicker and include both casual thick twins and thin lamellae. Increasing ER to 25:1 shortens twin strings and reduces their width and thickness. At ER 39:1, twins become thinner, more elongated, and form long strings, suggesting localized, periodic strain accommodation by micro-twins. Herein, Paramatmuni et al. [15] studied Schmid and non-Schmid twin behavior in Mg alloys, observing similar twins with lower frequency, greater elongation, and grain boundary constraints. The inhomogeneous distribution of twin strings, visible only in some coarse grains, is attributed to differences between the local and global Schmid factors of the parent grain.

Bimodal microstructures are preferred over equiaxed fine-grained structures for further straining [16]. Twins serve as preferential sites for recrystallization nuclei, making their characteristics crucial alongside grain size distribution in selecting thermomechanical processing routes. Twins indicate lower Schmid factors, signifying higher strain accommodation capacity and reduced crack initiation or propagation risk. To reduce experimental variation, ER 39:1, meeting both preconditions of a bimodal structure with desirable twin morphology and dispersion, was selected for upcoming hot rolling experiments.

**Hot Rolling:** Fig. 2 presents the microstructure of as-rolled samples at different annealing times and rolling speeds (26- and 130 mm s$^{-1}$). To preserve the initial microstructure for hot-rolling experiments, extruded samples (RE 39:1) were annealed at the hot-rolling temperature for two soaking durations, as shown in Fig. 2(a-b). Statistical analysis of grain size distribution and twin area fraction is summarized in Table II. Fig. 2(a-b) and Table II show that prolonged annealing increases microstructural inhomogeneity, evidenced by higher CV values. The microstructure includes small recrystallized grains and larger grains with a high twin area fraction. Extended annealing results in a bimodal structure with reduced twin fractions, attributed to texture randomization and detwinning during prolonged annealing [17].

As discussed in the hot extrusion section, twins in the microstructure lower the Schmid factor, facilitate further straining, and act as nucleation sites for recrystallization. Deformation with a 2.5-minute soaking time between passes produces smaller grains due to the presence of twins and higher stored energy, as seen in Fig. 2(a1-a2), compared to the 10-minute soaking time in Fig. 2(b1-b2). Longer annealing in pure Mg allows more dislocation recovery and rearrangement, reducing dislocation density and the internal stress needed for twin formation.

The rolling speed influences the strain rate during hot rolling, affecting dynamic recrystallization (DRX) grain size, as described by the Zener-Holloman parameter (($D_{DRX} = AZ^{-p}$), where p ranges from 0.1 to 0.35 and A is a material constant). Higher strain rates lower the critical strain for DRX, accelerate dislocation movement, and increase the Zener-Holloman parameter, resulting in finer DRX grains [18]. This is evident in Fig. 2, where higher rolling speeds (a2, b2) produce finer microstructures compared to lower speeds (a1, b1).

To optimize thermomechanical processing, mechanical properties like hardness must be assessed. Fig. 3 shows Vickers hardness measurements of hot-rolled wires, performed perpendicular to the rolling direction to evaluate strain homogeneity radially. This approach ensures microstructural uniformity, crucial for subsequent drawing deformability [13]. Fig. 3(a) shows that as strain increases, dynamic recrystallization (DRX) intensifies, straining many recrystallized grains and raising hardness. Higher strain rates increase DRX grain fraction and refine grains, contributing to strain-free structures. However, according to the Zener–Hollomon relation, higher strain rates also promote softening, reducing hardness values.

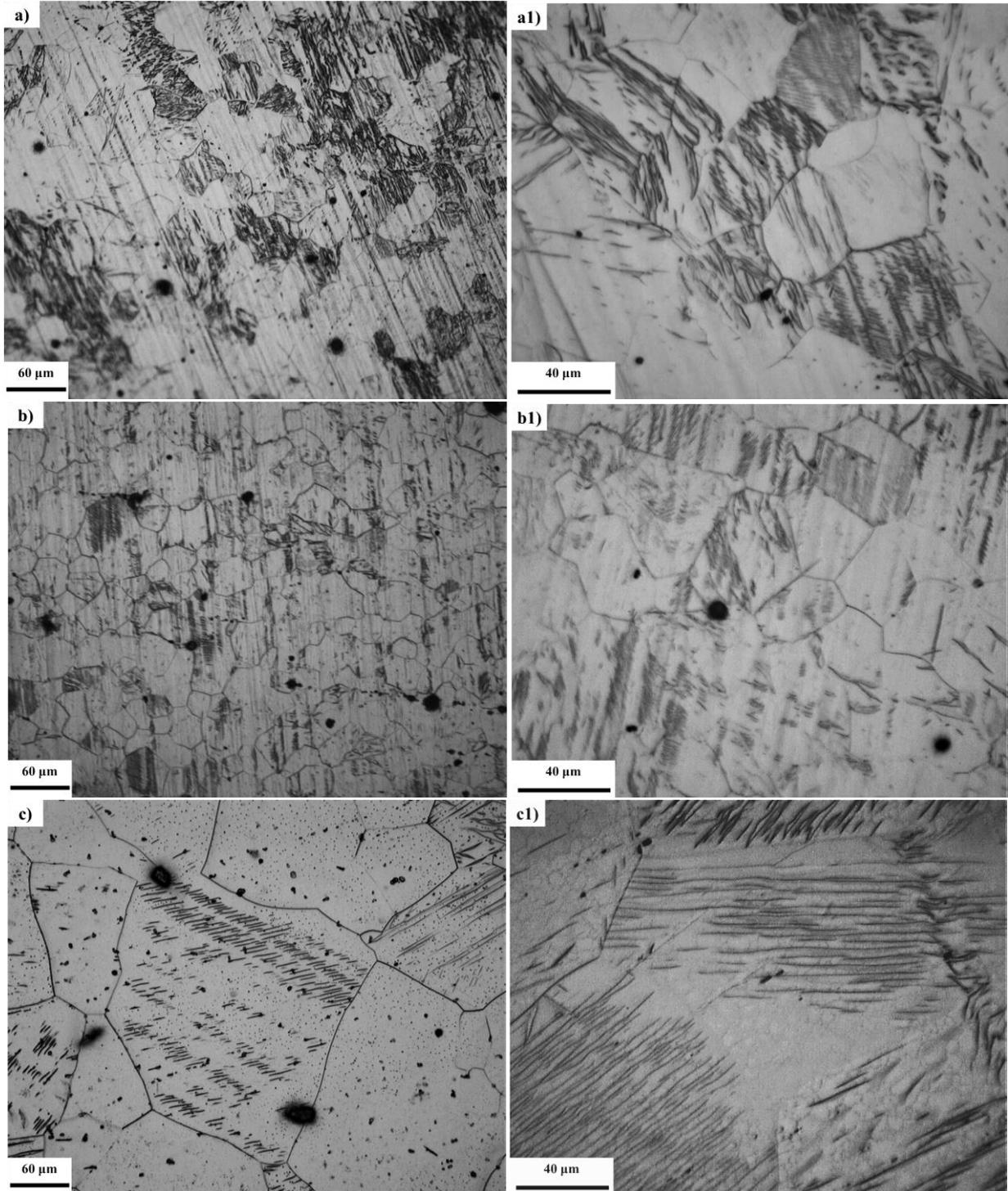

Fig. 1: Optical micrographs of as-extruded samples at different extrusion ratios: (a-a1) 6:1, (b-b1) 25:1 and (c-c1) 39:1.

TABLE I: THE STATISTICAL RESULTS OF AS-EXTRUDED PURE MG BILLETS AT DIFFERENT EXTRUSION RATIOS

| Extrusion Ratio | Average Grain Size (µm) | Standard Deviation | Coefficient of Variation (CV) | Grain Size Distribution |
|---|---|---|---|---|
| 6:1 | 35.71 | 21.24 | 0.595 | 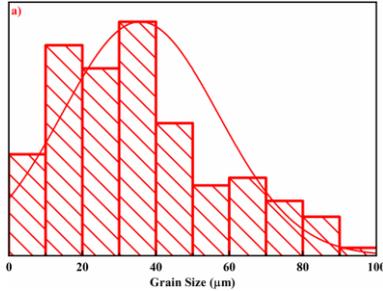 |
| 25:1 | 28.19 | 12.83 | 0.455 | 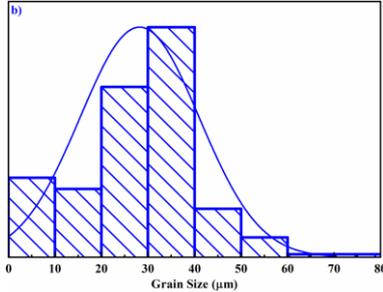 |
| 39:1 | 52.73 | 40.43 | 0.7655 | 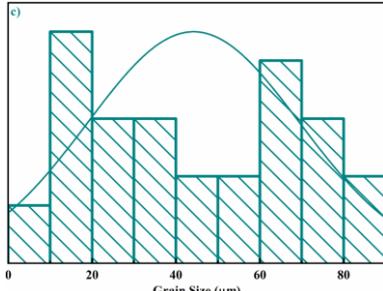 |

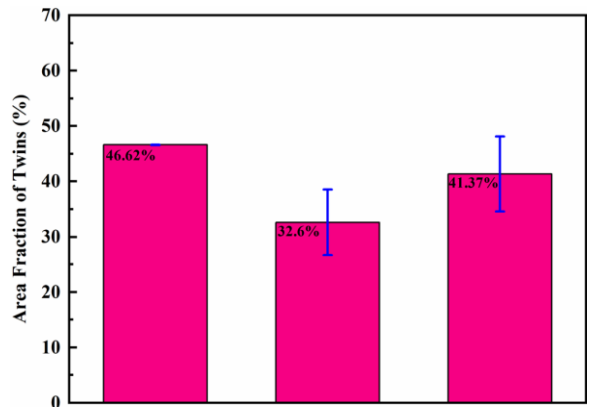
Area fraction of twins, for ER (a) 6:1, (b) 25:1, (c) 39:1

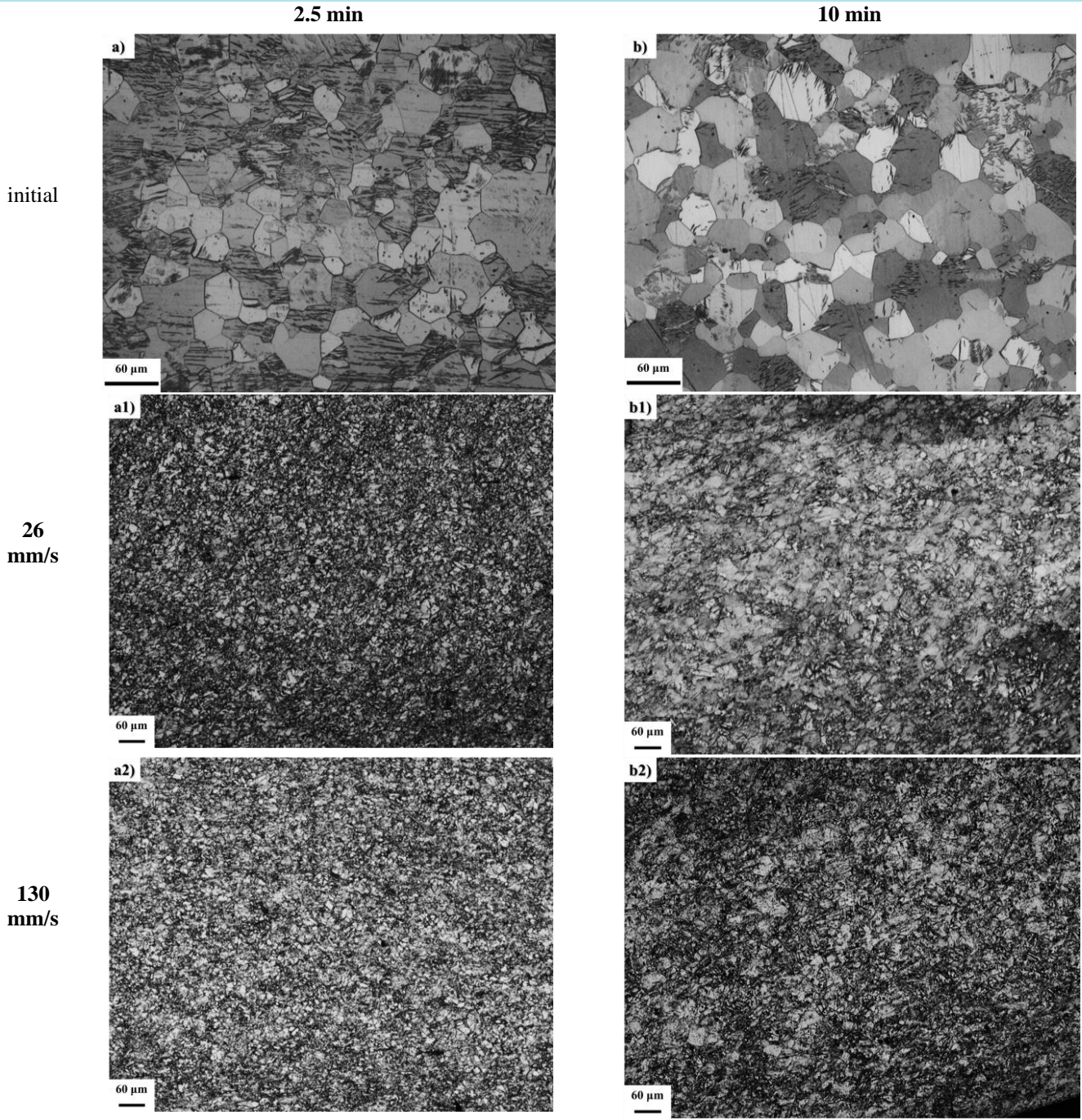

Fig. 2: Optical micrographs of the rolling experiment: a (a1-a2) and b (b1-b2) are related to annealed as-extruded rods at 350 °C for 2.5 and 10 minutes, respectively, and considered as initial microstructure of hot rolling. a1-b1 and a2-b2 are results of rolling at 26- and 130 mm s$^{-1}$ respectively.

TABLE II: THE STATISTICAL RESULTS OF AS-ANNEALED PURE MG RODS AT DIFFERENT ANNEALING TIMES

| Soaking Time (min) | Average Grain Size (μm) | Standard Deviation | Coefficient of Variation (CV) | Grain Size Distribution |
|---|---|---|---|---|
| 2.5 | 30.637 | 13.173 | 0.43 | |
| 10 | 37.029 | 21.396 | 0.578 | |

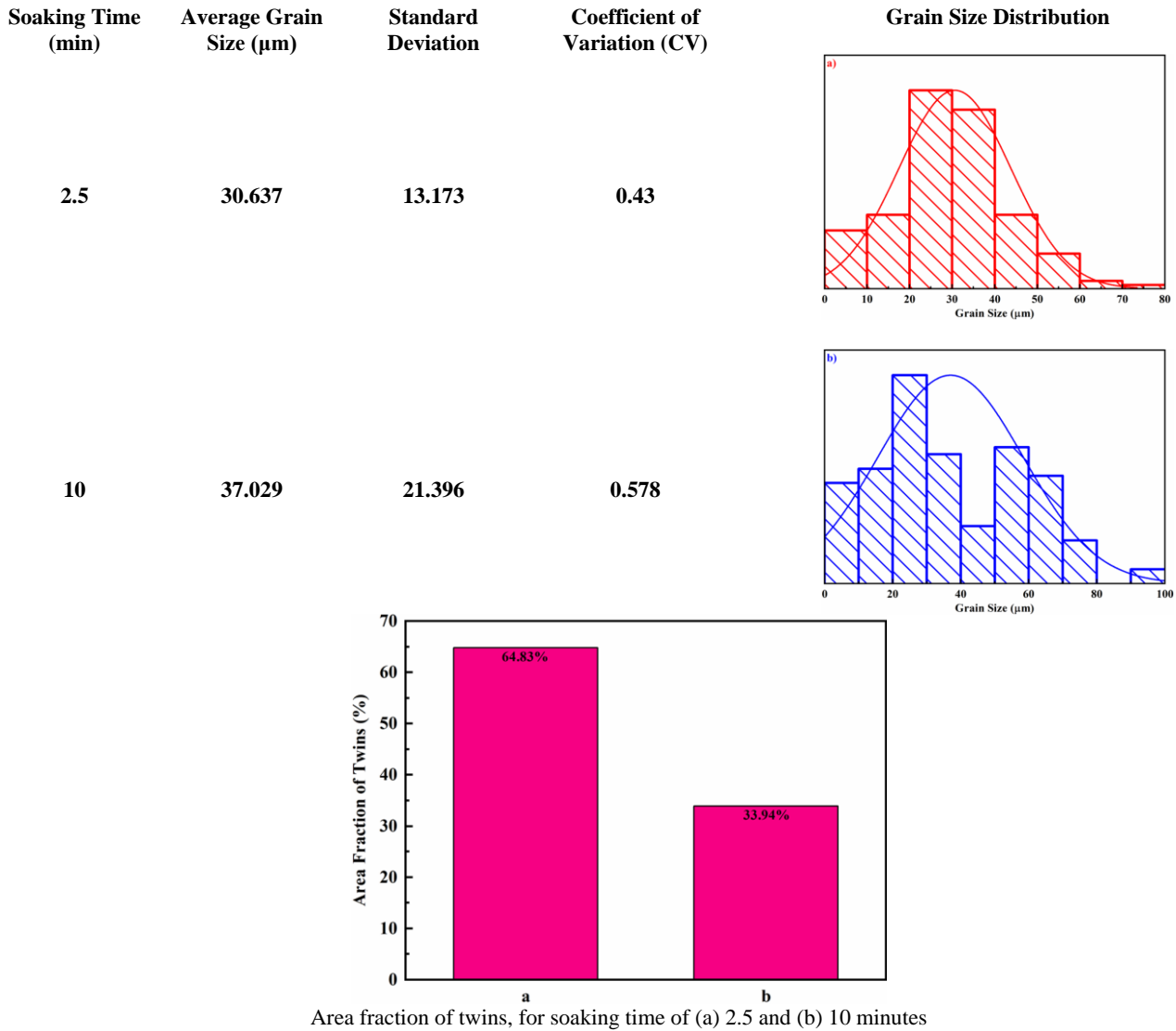

Area fraction of twins, for soaking time of (a) 2.5 and (b) 10 minutes

Fig. 3(b) shows that increased soaking time reduces twin fractions, limiting DRX nucleation sites and strengthening the basal texture in Mg, which raises hardness but reduces ductility. At lower strain rates, non-uniform strain causes significant hardness variations, while higher strain rates result in more uniform hardness due to extensive DRX. In the 10 min-26 mm s$^{-1}$ sample, prolonged annealing promotes recovery and substructure evolution, increasing hardness. Overall, shorter soaking times yield higher average hardness.

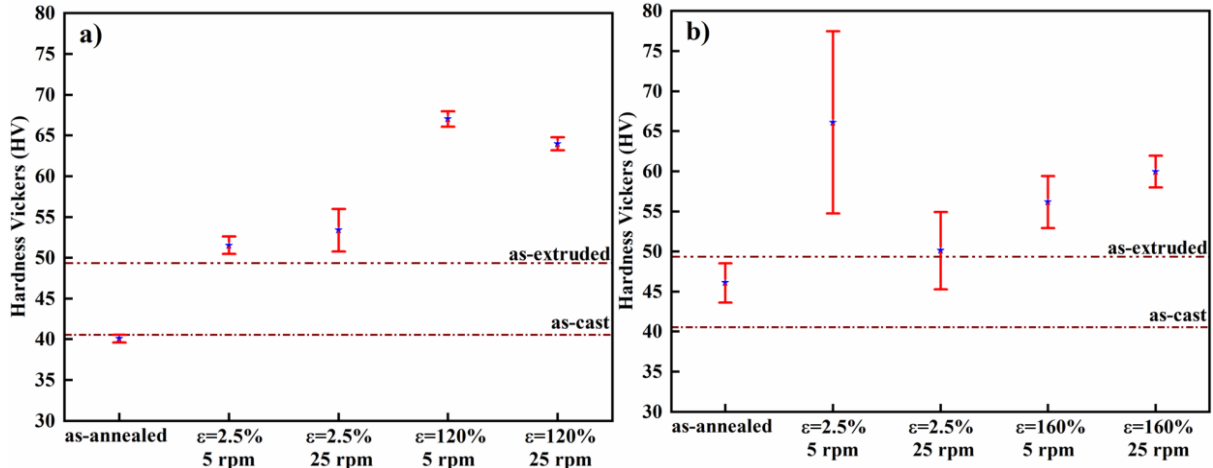

Fig. 3: Hardness distribution of samples related to interpass annealing times of (a) 2.5 and (b) 10 minutes

To minimize experiments, the optimal condition for cold drawing requires low hardness for sufficient plasticity and formability. The sample hot-rolled at 26 mm s$^{-1}$ with a 10-minute interpass annealing time best meets these criteria, showing superior performance in the cold drawing process.

**Cold Drawing:**
Fig. 4 shows the initial (a-a1), cold-drawn (b-b1), and annealed (c-c1) microstructures from the wire drawing experiment. The annealed samples (c-c1) underwent 5-minute annealing at 275°C, with other conditions assessed through hardness measurements. The initial microstructure (Fig. 4(a-a1)) consists of equiaxed recrystallized grains with a high twin area fraction, as shown in Table III. After ~12% logarithmic strain during cold deformation, the high SFE of pure Mg likely promoted dynamic recovery at grain boundaries, inhibiting global DRX and forming thick, elongated twins, contributing to the increased twin fraction in the as-drawn sample [19].

After deformation, samples were annealed at various times and temperatures, with the microstructure of the 275°C-5 minute sample analyzed (Fig. 4(c-c1)). The annealing formed limited strain-free recrystallized grains, while most grains remained un-recrystallized. The treatment transformed twins into thin, elongated structures. Further cold drawing of this sample led to fracture, as detailed in the hardness analysis. Fig. 5 shows the hardness of samples annealed at 275°C and 375°C. At 275°C, a 2.5-minute soaking time results in hardness similar to the cold-drawn sample, indicating recovery as the dominant process. At 5 minutes, partial recrystallization reduces hardness, while 7.5 minutes increases recrystallized grain fraction, further lowering hardness. The 10-minute sample shows similar hardness to the 7.5-minute sample, suggesting full recrystallization and a steady state by 7.5 minutes. In Fig. 5(b), the hardness of the sample annealed at 375°C for 2.5 minutes drops significantly due to faster recrystallization kinetics from enhanced dislocation climb and cross-slip. Beyond 2.5 minutes, the hardness shows a negligible increase, possibly due to the Hall-Petch effect from annealing twin formation. The optimal annealing condition for cold-drawn wires is 2.5–5 minutes at 375°C, yielding a soft, ductile wire suitable for further cold working.

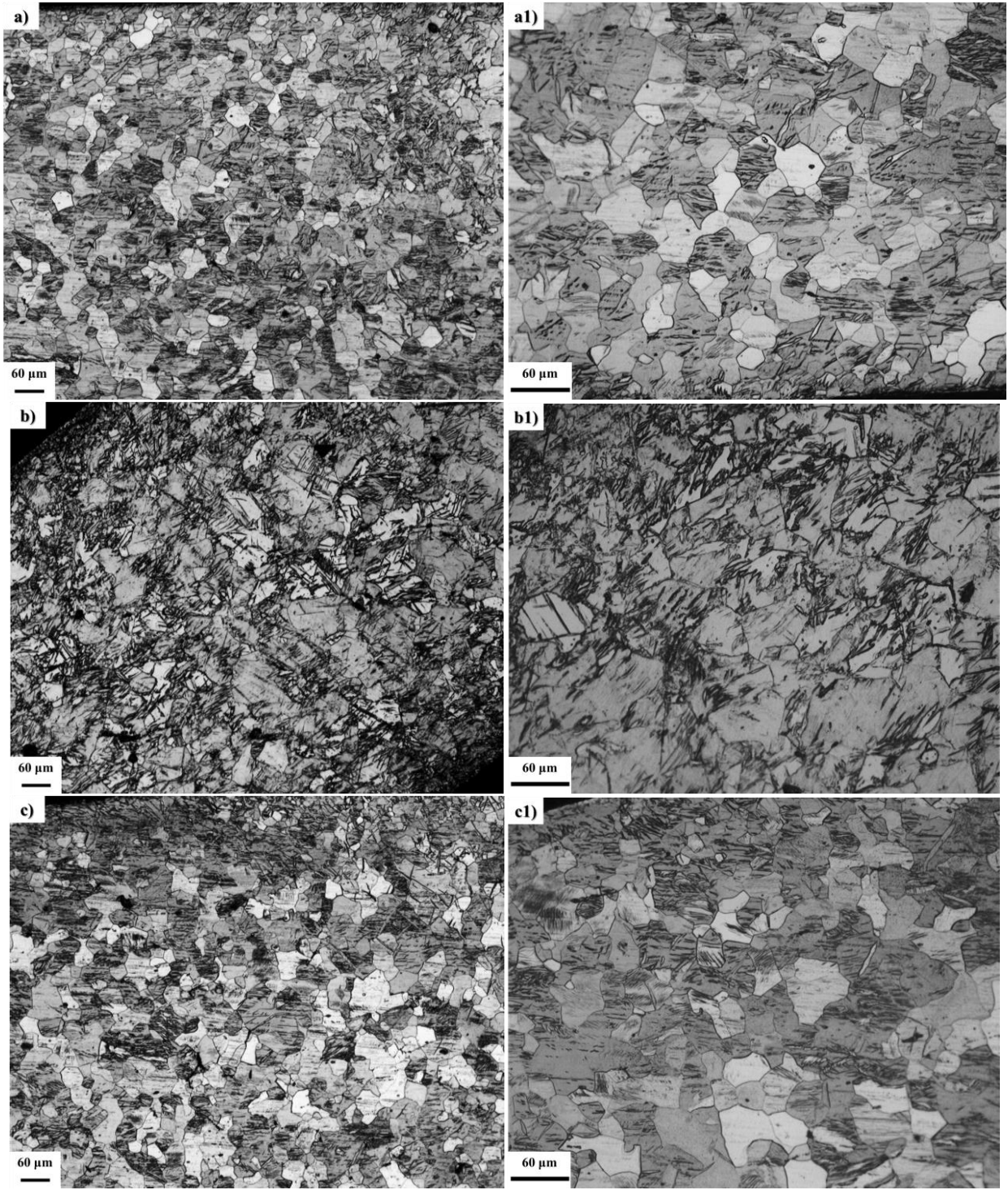

Fig. 4: Optical micrographs of the cold-drawn samples, (a-a1) initial microstructure of the deformation, (b-b1) as-drawn samples which were drawn with a logarithmic strain of 12.2%, (c-c1) annealed microstructure of deformed wires, tempered at 275 °C for 5 minutes.

TABLE III: THE STATISTICAL RESULTS OF COLD DRAWN WIRES

| State | Average Grain Size (μm) | Standard Deviation | Coefficient of Variation (CV) | Grain Size Distribution |
|---|---|---|---|---|
| annealed - prior to the cold drawing | 29.22 | 11.79 | 0.403 | a) |
| as-drawn | 36.24 | 17.92 | 0.494 | b) |
| annealed - after cold drawing | 32.24 | 13.02 | 0.404 | c) |

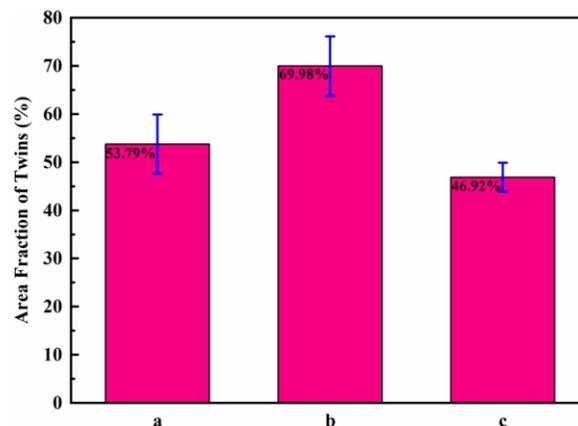

Area fraction of twins, (a) prior to the drawing, (b) as-drawn, (c) annealed after drawing

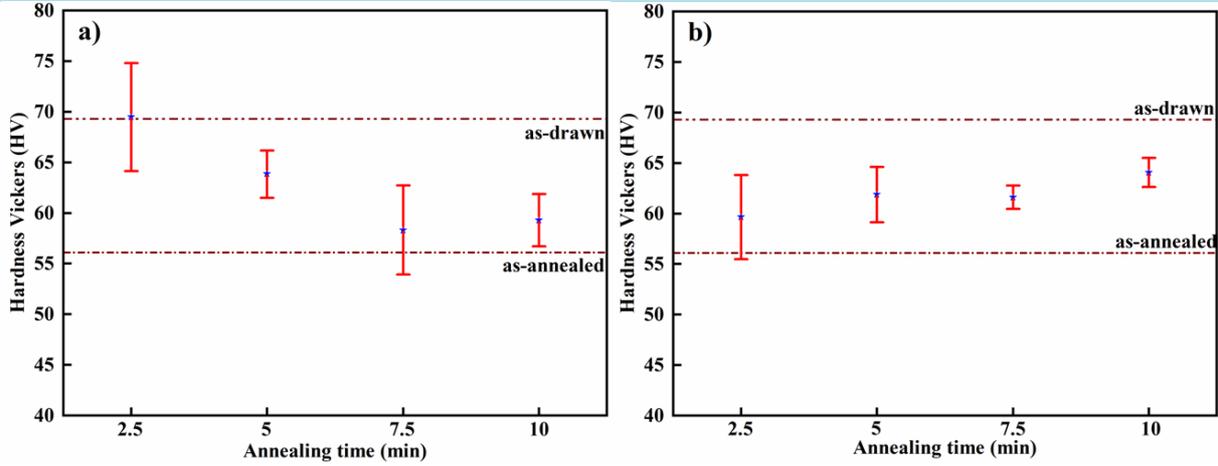

Fig. 5: Hardness distribution of samples, (a) interpass annealing at 275 °C, and (b) interpass annealing at 375 °C.

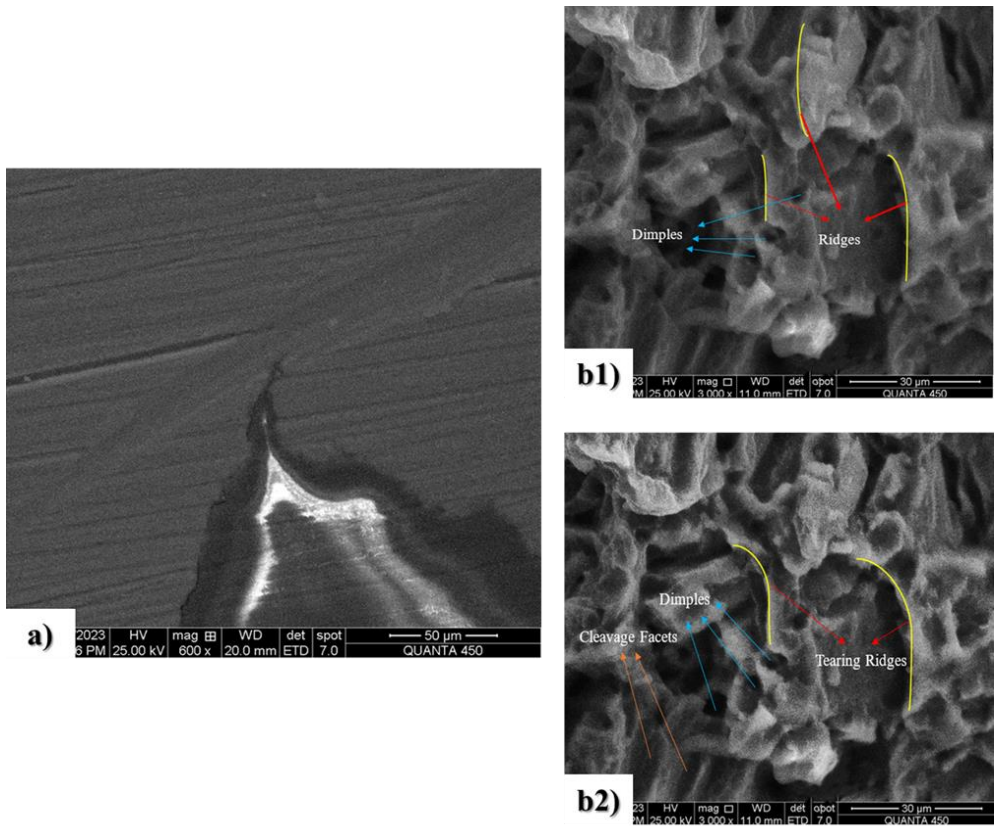

Fig. 6: (a) SEM micrograph of crack tip, at the center of the wire with annealing condition of 275 °C-5 min, (b1-b2) fracture morphology of the broken wire

Fracture morphology of the broken wire investigated utilizing SEM images as shown in Fig. 6. When a propagating twin interacts with a grain boundary, high stress concentration can develop in the adjacent grain, potentially leading to twin nucleation or crack initiation. A running cleavage crack interacting with a twin may change its path to an equivalent habit system within the twin or along the twin-matrix interface, as observed in Be and Mg alloys. Interaction between a cleavage crack and a grain boundary can result in bifurcation or a transition from trans-granular to intergranular fracture [20].

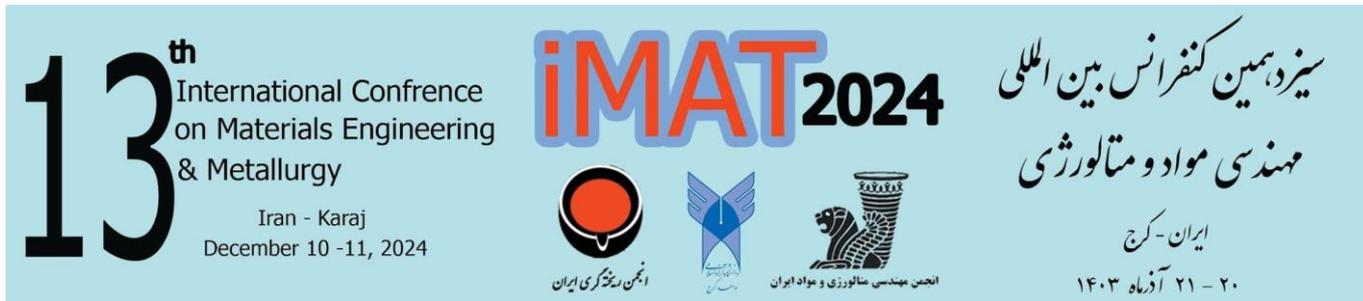

The crack grows along the twin–matrix interface, the preferred path due to strain incompatibility and high stress concentrations. In Fe–Si single crystals, slip steps at the twin–matrix interface from dislocation dissociation creates weaknesses that increase stress concentration [21, 22]. As the crack propagates (Fig. 6(a)), it may blunt due to twin intersection and the curved plastic zone forming ahead of the crack tip.

SEM images of the fracture surface (Fig. 6(b1-b2)) show both ductile and brittle features in magnesium under tensile loading. Tearing ridges indicate ductile fracture, while deep dimples, formed at 275°C, reflect enhanced deformation. Cleavage facets suggest brittle behavior. Overall, the wire exhibits predominantly ductile behavior with minor brittle features, likely from twin-induced fractures, indicating drawability is more influenced by twin characteristics than restored plasticity.

## IV. Conclusion

The systematic investigation of thermomechanical processing for pure Mg yielded several significant findings across multiple processing stages:

In hot extrusion, an extrusion ratio of 39:1 at 350°C produced the optimal microstructure, characterized by:
- A beneficial bimodal grain distribution
- Ultra-thin twin lamellae with periodic behavior
- Enhanced strain accommodation capability due to the combination of fine and coarse grains

During hot rolling, the following key relationships were established:
- Higher rolling speeds (130 mm $s^{-1}$) resulted in finer microstructure due to increased dynamic recrystallization
- Shorter interpass annealing times (2.5 minutes) led to smaller grains and higher twin fraction
- Samples rolled with 26 mm $s^{-1}$ speed and 10 minutes interpass annealing time showed optimal formability for subsequent drawing

For cold drawing operations:
- The optimal annealing condition was determined to be 375°C for 2.5-5 minutes, providing sufficient softening for further deformation
- Twin characteristics proved more critical than restored plasticity in determining wire drawability
- Fracture analysis revealed a mixed ductile-brittle behavior, with twin-matrix interfaces serving as preferential crack propagation paths

This comprehensive study establishes a practical processing route for pure magnesium wire production, demonstrating that careful control of processing parameters and microstructural evolution is crucial for achieving desired formability. The findings highlight the critical role of twin characteristics, grain size distribution, and restoration processes in determining the success of multi-step thermomechanical processing of pure magnesium.

These results provide valuable insights for the industrial processing of pure magnesium and establish a foundation for further optimization of magnesium processing parameters for specific applications.